\documentclass[11pt,a4paper]{article}
\usepackage[utf8]{inputenc}
\usepackage{amsmath}
\usepackage{amsthm}
\numberwithin{equation}{section}
\usepackage{amsfonts}
\usepackage{mathtools}
\usepackage{mathrsfs}
\usepackage{mathpazo}
\usepackage{multirow}
\usepackage{amssymb}
\usepackage{graphicx}
\usepackage{ifpdf}
\usepackage{braket}
\usepackage[dvipsnames]{xcolor}

\usepackage{cite}
\usepackage[bookmarks=true,colorlinks=true,linkcolor=black,citecolor=orange,urlcolor=orange,bookmarksnumbered]{hyperref}

\usepackage[left=2cm, right=2cm, top=2cm, bottom=3cm]{geometry}
\begin{document}

\begin{center}

\vspace{1.0truecm}

{\Large \bf On the universal exceptional structure 

of world-volume theories in string and M-theory}

\vspace{1.0truecm}

{David Osten}

\vspace{0.5truecm}

{\em Institute for Theoretical Physics (IFT), \\
University of Wroc\l aw \\
pl. Maxa Borna 9, 50-204 Wroc\l aw, Poland
}

\vspace{0.2truecm}

{{\tt david.osten@uwr.edu.pl}}

\vspace{0.2truecm}
\end{center}

\begin{abstract}
A universal structure of world-volume theories of $\frac{1}{2}$-BPS branes in string and M-theory in terms of exceptional generalised geometry is observed. 
Previous constructions are extended in two ways: from internal $d$-dimensional space to full 11- or 10-dimensional spacetimes, by coupling to the tensor hierarchy gauge fields, and from $E_{d(d)}$ with $d\leq 6$ to $E_{7(7)}$ and $E_{8(8)}$. This is done by a clarification of the role of the tensor hierarchy and its end in the context of brane world-volume theories. 
The exceptional structure of the gauged $\sigma$-model of the Kaluza-Klein monopole is provided as a new example.
\end{abstract}

\vspace*{0.5cm}

\section{Introduction and general setting}
Symmetry principles are at the core of modern approaches to high energy physics. In this context, the program of generalised geometries and extended field theories aims to promote $T$- and $U$-dualities of string and $M$-theory to proper symmetries related to the duality groups O$(d,d)$ or $E_{d(d)}$. Remarkably, it turned out that both string theory and supergravity, as its low energy effective action, possess some underlying structure associated to these duality groups O$(d,d)$ resp. $E_{d(d)}$ -- even if the backgrounds considered do not permit such dualities. For example, a central statement of exceptional field theory \cite{Hull:2007zu,Pacheco:2008ps,Berman:2010is,Berman:2011jh,Coimbra:2011nw,Coimbra:2011ky,Berman:2012vc,Berman:2012uy,Coimbra:2012af,Hohm:2013pua,Godazgar:2013dma,Hohm:2013vpa,Hohm:2013uia,Hohm:2014fxa,Lee:2014mla,Hohm:2014qga,Musaev:2015ces,Abzalov:2015ega,Cederwall:2015ica,Berman:2020tqn,Hulik:2023aks} is that maximal supergravity can be phrased in an $E_{d(d)}$-covariant way. In this article, we present more conclusive evidence about an accompanying statement for world-volume theories. In particular, we consider the predestined $\frac{1}{2}$-BPS branes -- so those whose supergravity solutions preserve half of the supersymmetries:
\begin{center}
\textit{The classical world-volume theories of $\frac{1}{2}$-BPS branes in string and M-theory have a universal Hamiltonian description in terms of generalised exceptional geometry.}
\end{center}
The connection of world-volume theories and duality symmetries goes back to \cite{Duff:1989tf,Tseytlin:1990nb,Tseytlin:1990va,Siegel:1993bj,Siegel:1993th} where the string world-volume was phrased in an O$(d,d)$-covariant way. Both approaches were developed further in several branches of research -- so called $\mathcal{E}$-models that are useful for generalised dualities and the study of integrable models \cite{Klimcik:1995ux,Sfetsos:1999cc,Demulder:2018lmj,Hassler:2020xyj,Borsato:2021gma,Borsato:2021vfy}, or the connection to non-commutative geometry \cite{Blair:2014kla,Osten:2019ayq}. This framework can also be extended to heterotic string where the classical phase space can be shown to be O$(d,d+n)$-covariant \cite{Hatsuda:2022zpi,Osten:2023cza}.

The literature on world-volume theories of higher dimensional $P$-brane objects, like the D- or M-branes, in string and M-theory in the context of exceptional field theory is sparse -- mainly due to the fact that their role in the context of M-theory is not as clear as the one of the string in string theory. Nevertheless, since the appearance of exceptional field theory and exceptional generalised geometry several advances into reformulations of the M2-brane \cite{Duff:1990hn,Duff:2015jka,Hatsuda:2012vm,Strickland-Constable:2021afa}, the M5-brane \cite{Hatsuda:2013dya}, D-branes \cite{Hatsuda:2012uk}, as well as general approaches to formulate exceptionally covariant actions \cite{Sakatani:2016sko,Blair:2017hhy,Sakatani:2017vbd,Arvanitakis:2018hfn,Blair:2019tww,Sakatani:2020umt,Hatsuda:2023dwx}. Related algebraic structures also appeared on the level of QP-manifolds \cite{Arvanitakis:2018cyo,Arvanitakis:2021wkt,Arvanitakis:2022fvv,Osten:2023iwc}. 

In these previous works, a pattern emerged that was explored in detail in \cite{Osten:2021fil}. Let us introduce this result here, as it and its shortcomings are the starting point for this article. 

\subsection{Input from generalised geometry}
Central input is the duality group $G$ together with the so-called tensor hierarchy of representations $\mathcal{R}_1,\mathcal{R}_2, ...$ of $G$ together with two operations, $\bullet: \ \mathcal{R}_p \times \mathcal{R}_q \rightarrow \mathcal{R}_{p+q}$ and $\partial : \ \mathcal{R}_p \rightarrow \mathcal{R}_{p-1}$. Here, these will be expressed as follows:
\begin{align}
(A \bullet B)^{M_{p+q}} &= {\eta^{M_{p+q}}}_{K_p L_q} A^{K_p} B^{L_q}  \label{eq:TensorHierarchy}, \quad (\partial A)^{M_{p-1}} = {D_{M_p}}^{M_{p-1},L} \partial_L A^{M_p}
\end{align}
for  $A\in \mathcal{R}_p, B \in \mathcal{R}_q$. $K_p,L_p,M_p,...$ are the indices of $\mathcal{R}_p$. As a short-hand notation, we use $K,L,M,...$ as $\mathcal{R}_1$-indices and $\mathcal{K},\mathcal{L},\mathcal{M}...$ for $\mathcal{R}_2$, if only these two representations are relevant. 

For the typical duality groups O$(d,d)$, O$(d,d+n)$ and $E_{d(d)}$ the $\eta$- and $D$-symbols are given explicitly in a case by case analysis \cite{Hohm:2013vpa,Hohm:2013uia,Hohm:2014fxa,Abzalov:2015ega,Musaev:2015ces,Hohm:2015xna,Wang:2015hca}, or as coming from some underlying structure \cite{Palmkvist:2013vya,Cederwall:2018aab,Cederwall:2019bai,Cederwall:2019qnw}. More generally, this structure fits into the structure of a differential graded Lie algebra, see \cite{Bonezzi:2019ygf,Lavau:2019oja,Bonezzi:2019bek,Arvanitakis:2018cyo,Osten:2023iwc}. These imply several consistency conditions, which are related to the (graded) Jacobi identities of the underlying differential graded Lie algebra.

For the world-volume theories the main role is played by the representations $\mathcal{R}_1$ and $\mathcal{R}_2$, the corresponding $\eta$-symbol $\eta_{\mathcal{L},MN}$: $\mathcal{R}_1 \otimes \mathcal{R}_1 \rightarrow \mathcal{R}_2$ and the derivation $\partial: \mathcal{R}_2 \rightarrow \mathcal{R}_1$. Decomposed into GL$(d)$-indices $m,n,...$ --- the co-called M-theory section -- these representations are
\begin{align}
	\mathcal{R}_1: \quad V^M &= (V^m , V_{m_1 m_2} , V_{m_1 ... m_5} , V_{m,m_1 ... m_7}, ...) \nonumber \\
	\mathcal{R}_2: \quad W^\mathcal{M}&= (W^m, W^{m_1...m_4} , W^{m,m_1...m_6},...) \nonumber
\end{align}
with additional terms appearing at $d \geq 8$. Explicit expressions for this relevant $\eta$-symbol, also for the IIb-section, can be found in \cite{Sakatani:2017xcn}.

These representations are associated to certain bundles which can be read off the above index decomposition. The $\mathcal{R}_1$-representation is associated to the so-called \textit{generalised tangent bundle}, \linebreak $(T \oplus \bigwedge^q T^\star \oplus ...) M$, on which one can define an $E_{d(d)}$-covariant \textit{generalised Lie derivative}
\begin{equation}
\mathcal{L}_{\Phi_1} \Phi_2^N = \Phi_1^M \partial_M \Phi_2^N - \Phi_2^M \partial_M \Phi_1^N + {Y^{MN}}_{KL} (\partial_M \Phi_1^K) \Phi_2^L
\end{equation}
for sections $\Phi_i$ of the generalised tangent bundle. In this article, we will use the following definition of the so-called $Y$-tensor:
\begin{equation}
	{Y^{KL}}_{MN} = \eta_{\mathcal{P},MN} D^{\mathcal{P}L,K}. \label{eq:SectionCondition}
\end{equation}
This only superficially differs from the more standard definitions, for example in \cite{Berman:2012vc,Sakatani:2017xcn,Berman:2020tqn}, but allows to directly generalise certain results to $E_{7(7)}$ and $E_{8(8)}$. This $D$-symbol corresponds to the derivation $\mathcal{R}_2 \rightarrow \mathcal{R}_1$ and coincides with the $\eta$-symbol $\eta^{\mathcal{P},KL}$ for $d\leq 6$.

\paragraph{Section condition.} In order to obtain manifest $E_{d(d)}$-covariance, we introduce generalised coordinates $X^M = (x^m,...)$. Physical functions $f,g$ are subject to the section condition:
\begin{equation}
	\eta^{\mathcal{L},MN} \partial_M f \partial_N g =0
\end{equation}
which ensures that they only depend on the physical coordinates $x^m$. For $E_{d(d)}$, there are two solutions: the $d$-dimensional M-theory section and the $(d-1)$-dimensional type IIb solution. The indices $k,l,m,...$ are reserved for these GL$(d)$- resp. GL$(d-1)$-indices.

\paragraph{Geometry.} In order to describe a full 11- or 10-dimensional supergravity background, we take coordinates $Y = (y^\mu , X^M)$. There $X^M$ are the generalised coordinates of the $d$- or $(d-1)$-dimensional \textit{internal} geometry and $\kappa,\lambda,\mu,...$ are indices for the $($11/10$-d)$-dimensional \textit{external} coordinates $y^\mu$.

The internal geometry is described by the so-called generalised metric $\mathcal{H}^{MN} \in \frac{E_{d(d)}}{H_d}$, where $H_d$ is the maximal compact subgroup of $E_{d(d)}$. This generalised metric encodes both the internal metric and internal $p$-form gauge fields. For the complete $11$- or $10$-dimensional background one additionally needs a $($11/10$-d)$-dimensional metric $g_{\mu\nu}$ and a hierarchy of $p$-form gauge fields $A^{(p)}_{\mu_1 ,... \mu_p} \in \mathcal{R}_p$ for $p = 1, ... , (8-d)$. The general construction of exceptional field theory for $d \leq 8$ can be found in \cite{Berman:2020tqn}. 

\subsection{The existing setting}
It had been shown that the (single) world-volume dynamics of strings, lower dimensional D-branes and NS5-branes (in the type IIb section), and M2- and M5-branes (in the M-theory section), in the \textit{internal space} of the exceptional field theory, can all be described by the following unifying construction for $P$-branes. 
\begin{itemize}
\item The \textit{phase space variables} are currents $z_M \in \mathcal{R}_1$, which are (spatial) $P$-forms, i.e. spatial volume forms, on a $P$-brane world-volume. Their prototypical form (in the corresponding section) is 
\begin{equation}
z_M = (p_\mu , ... , \mathrm{d} x^{\mu_1} \wedge ... \wedge \mathrm{d} x^{\mu_P})
\end{equation}
for a $P$-brane. $p_\mu$ denotes the canonical momentum field and the dots $...$ correspond to potential additional world-volume fields on the $P$-brane. See \cite{Osten:2021fil} for examples.

\item The \textit{current algebra}, which encodes the Poisson brackets of the currents,
\begin{equation}
\{z_M(\sigma) , z_N(\sigma^\prime) \} = \eta_{\mathcal{L},MN} \mathcal{Q}^\mathcal{L} \wedge \mathrm{d}\delta(\sigma- \sigma^\prime) \label{eq:IntroCurrentAlgebra}
\end{equation}
is characterised by some spatial $(P-1)$-form $\mathcal{Q}^\mathcal{M} \in \mathcal{R}_2$ and the $\eta$-symbols.

\item For \textit{generalised covariance} of such a current algebra, and hence also the full model, we impose that fields $\Phi_i = \int \Phi^M_i (X(\sigma)) z_M(\sigma)$, that are associated to sections of the generalised tangent bundle, satisfy
\begin{equation}
	\{ \Phi_2 , \Phi_1 \} = \mathcal{L}_{\Phi_1} \Phi_2. \label{eq:CurrentGeneralisedLie}
\end{equation}
This leads to the so-called \textit{brane charge constraints}: $\mathcal{Q}^\mathcal{M} \wedge \mathrm{d} X^N \partial_N = \eta^{\mathcal{M},KL} z_K \partial_L $. Such 'brane charges' had been discussed in similar forms in the construction of actions in \cite{Sakatani:2016sko,Blair:2017hhy,Sakatani:2017vbd,Arvanitakis:2018hfn,Blair:2019tww,Sakatani:2020umt}. In this exact form they appeared first in \cite{Osten:2021fil}.

\item The \textit{Hamiltonian} and \textit{spatial diffeomorphism} constraints are characterised by the generalised metric and the $\eta$-symbol\footnote{The integral domain is a spatial slice of the world-volume, parametrised by the spatial coordinates $(\sigma_1,...,\sigma_P)$. 
}
\begin{align}
	H &= \frac{1}{2} \int \mathcal{H}^{MN} z_M z_N \label{eq:Hamiltonian}, \qquad 0 = \eta^{\mathcal{M},KL} z_K z_L.
\end{align}
\end{itemize}

\subsection{Limits and new insights} \label{chap:Limits}
The limits of the above construction are two-fold so far:
\begin{itemize}
	\item It focusses only on the $d$-dimensional internal space of $E_{d(d)}$ exceptional field theory. 
	\item These results are restricted to $d \leq 6$ with according restrictions on the dimension of the branes that could be described by this formalism. This restriction is the limiting factor for potential prediction of world-volume theories of exotic branes.
\end{itemize}
The present article attacks both of these problems. The solution consists of two main steps:
\begin{enumerate}
	\item an extension of the current algebra with currents $t^{(p)}_{M_p} \in \mathcal{R}_p$ that correspond to generators of the tensor hierarchy. This will allow a coupling of the world-volume to the tensor hierarchy gauge fields $A^{(p)}_{\mu_1 ... \mu_p}$ in sections \ref{chap:THCurrents} and \ref{chap:ExtSpace}.
	\item a slight modification in comparison to approaches in exceptional field theory at the end of the tensor hierarchy. The tensor hierarchy that is used here is the one that appears for example in Table 1 in \cite{West:2018lfn} and was derived from the $E_{11}$-approach \cite{West:2003fc,West:2004iz,West:2004kb,Tumanov:2016abm,Bossard:2021ebg}. This includes representations up to $\mathcal{R}_{11-d}$. Typically, the relevant representations -- and their associated $p$-form gauge fields -- for target space exceptional field theory end at $\mathcal{R}_{9-d}$. Contributions from the higher representations are fixed by Bianchi identities or drop out of the exceptional field theory action \cite{Hohm:2013vpa,Hohm:2013uia,Hohm:2014fxa,Berman:2020tqn}.
	
	For the world-volume theories, it seems necessary to introduce the full tensor hierarchy up to $\mathcal{R}_{11-d}$. In order to compare the current algebra representation to the fields of target space exceptional field theory, these higher representations are actually not necessary. As also expressions for their $\bullet$-product and derivation $\partial$ have not appeared in the literature, they will not need to be introduced explicitly here. The only key differences of the conventions in this article to the typical literature on target space exceptional field theory are that:
	\begin{itemize}
		\item  $\mathcal{R}_{9-d} = \mathbf{1} \oplus \mathbf{adj}$.
		\item For $E_{8(8)}$,  the $\mathcal{R}_2$-representation is taken to be: $\mathcal{R}_2 = \mathbf{1} \oplus \mathbf{248} \oplus \mathbf{3875}$. The corresponding operations $\bullet$, $\partial$ of the tensor hierarchy are introduced in section \ref{chap:E8}.
	\end{itemize}
This will give a convenient unifying picture for compensator fields in section \ref{chap:EndOfHierarchy}.
\end{enumerate}
With this, one can also describe the world-volume models for the $E_{7(7)}$- and $E_{8(8)}$-exceptional field theories. This is done in sections \ref{chap:E7} and \ref{chap:E8}. A new example of brane world-volume model for which this is necessary is the world-volume theory of the Kaluza-Klein monopole \cite{Bergshoeff:1997gy}. This is presented in the above formalism in section \ref{chap:KKM}.

\section{Tensor hierarchy currents and the brane charge constraints} \label{chap:THCurrents}

The current algebra, introduced above, already contains currents $z\in \mathcal{R}_1$ and $\mathcal{Q} \in \mathcal{R}_2$. This suggests that the full hierarchy of world-volume currents $t^{(p)}(\sigma) \in \mathcal{R}_p$ can be introduced on the phase space of a $P$-brane world-volume. These are spatial (world-volume) $(P-p+1)$-forms. Their (graded) Poisson structure is defined by the tensor hierarchy $\eta$-symbols \eqref{eq:TensorHierarchy}:
\begin{equation}
\{ t^{(p)}_{K_p}(\sigma) , t^{(q)}_{L_q}(\sigma^\prime) \} = {\eta^{M_{p+q}}}_{K_p L_q} t^{(p+q)}_{M_{p+q}} (\sigma) \wedge \mathrm{d}\delta(\sigma - \sigma^\prime). \label{eq:CurrentAlgebra}
\end{equation}

\paragraph{Brane charge constraints.} The generalised Lie derivative of two sections $\Lambda = \int \Lambda^M(x(\sigma)) t^{(1)}_M \in \mathcal{R}_1$, $\Phi = \int \Phi^{M_p}(x(\sigma)) t^{(p)}_{M_p} \in \mathcal{R}_p$ can be written as
\begin{equation}
	\mathcal{L}_{\Lambda} \Phi = \{ \Phi , \Lambda \} \label{eq:GeneralisedLieViaCurrent}
\end{equation}
if the currents $t^{(p)}$ satisfy the following hierarchy of \textit{brane charge constraints}:
\begin{equation}
	t_{K_{p+1}}^{(p+1)} \wedge \mathrm{d} X^L \partial_L = {D_{K_{p+1}}}^{M_p,L} t^{(p)}_{M_p} \partial_L . \label{eq:BraneChargeConstraint}
\end{equation}
As in the previous construction, solutions of these constraints will project out parts of the currents, that are relevant for the description of the typical brane models, from the abstract currents \eqref{eq:CurrentAlgebra}. In principle, this is a different representation of the recent construction on QP-manifolds \cite{Osten:2023iwc}. The remarkable difference is that the derivation $\partial$ is encoded by the $\delta$-function differential.

In order to derive the generalised Lie derivative \eqref{eq:GeneralisedLieViaCurrent} in the form $\mathcal{L}_\Lambda \Phi = \Lambda \bullet \partial \Phi + \partial (\Lambda \bullet \Phi)$ \cite{Wang:2015hca}, one needs the following identity
\begin{equation}
\left( {\eta^{L_p}}_{MN_p-1} { D_{K_p}}^{N_{p-1},Q} + {D_{N_{p+1}}}^{L_p,Q} {\eta^{N_{p+1}}}_{M K_p} \right) \partial_Q = \delta_{K_p}^{L_p} \partial_M . \nonumber
\end{equation}
When treating the tensor hierarchy as a differential graded Lie algebra or as QP-manifolds \cite{Arvanitakis:2018cyo,Osten:2023iwc}, this identity corresponds to a graded Jacobi identity $\{ \{ \Lambda , \Theta \} , \Phi \} = \{\Lambda , \{\Theta , \Phi\} \} + \{\Theta,\{\Lambda,\Phi\}\}$.

The space of solutions to the brane charge constraints is very restricted. For the $E_{d(d)}$-tensor hierarchies up to $d \leq 5$ these conditions have been solved in this form in \cite{Osten:2023iwc} and shown to correspond to the $\frac{1}{2}$-BPS branes. Issues appeared there for $d = 6$, as the tensor hierarchy was assumed to terminate at $\mathcal{R}_{9-d}$. Although this goes beyond the scope of this paper it seems plausible that the modified discussion of the end of the tensor hierarchy, mentioned in section \ref{chap:Limits}, will solve this problem and \eqref{eq:BraneChargeConstraint} will be sufficient conditions.

\paragraph{Brane charges via non-geometric coordinates.}  In \cite{Osten:2021fil}, a different approach had been proposed which only refers to the $\mathcal{R}_1$- and $\mathcal{R}_2$-representations. For this, a 'natural' ansatz for the $z_M$-current of a $P$-brane in terms of non-geometric coordinates needed to be made:
\begin{equation}
z_M = \frac{1}{P} \eta_{\mathcal{L},MN} \mathcal{Q}^{\mathcal{L}} \wedge \mathrm{d} X^N. \label{eq:CurrentNonGeometric}
\end{equation}
Together with \eqref{eq:GeneralisedLieViaCurrent} this gives a system of equations for $\mathcal{Q}$, the solutions of which correspond to the $\frac{1}{2}$-BPS branes. This approach works for world-volume theories with $d \leq 6$, so $P \leq 5$ or $6$. For example, it automatically gives the quite intricate structure of the M5 world-volume \cite{Pasti:1997gx}, that was first discussed in the context of exceptional generalised geometry in \cite{Hatsuda:2013dya}.

The condition \eqref{eq:CurrentNonGeometric} is ad-hoc, both due to appearance of non-geometric coordinates, for which the Poisson brackets are not known\footnote{In contrast to the string \cite{Blair:2014kla}.}, and due to the necessary factor $\frac{1}{P}$. With the full hierarchy of conditions \eqref{eq:BraneChargeConstraint} it seems that this additional assumption \eqref{eq:CurrentNonGeometric} is not necessary. Hence, we will not take use of this approach in the following.

\section{External space and tensor hierarchy gauge fields} \label{chap:ExtSpace}
The tensor hierarchy gauge fields in exceptional field theory are $p$-forms (in external space), valued in $\mathcal{R}_p$. They will be denoted by $A^{(1)} , A^{(2)} , ...$. The dynamics of a world-volume in a full 11- or 10-dimensional supergravity, with generalised coordinates $Y = (y^\mu , X^M)$ as introduced in the introduction, is proposed to be given by the Hamiltonian
\begin{equation}
H = \frac{1}{2} \int \left( \mathcal{H}^{MN} z_M z_N + g^{\mu\nu} \pi_\mu \pi_\nu \right). \label{eq:THHamiltonian}
\end{equation}
where the coupling to the tensor hierarchy gauge fields is encoded as follows
\begin{align}
	\pi_\mu &= p_\mu - A_\mu^{(1) M_1} t^{(1)}_{M_1} - A_{\mu \nu}^{(2) M_2} t^{(2)}_{M_2} \wedge \mathrm{d} y^\nu + ... \label{eq:THmomentum}
\end{align}
where $p_\mu$ is the associated momentum to the external coordinate field $y^\mu$. By using the brane charge condition \eqref{eq:BraneChargeConstraint}, one obtains
\begin{align}
&{} \quad \{ \pi_\nu(\sigma) , \pi_\mu(\sigma^\prime) \} \label{eq:CurrentTHexternal}
= - \left( F^{(1) M_1}_{\mu\nu} t^{(1)}_{M_1} (\sigma) + F^{(2) M_2}_{\mu \nu \lambda} t^{(2)}_{M_2} \wedge \mathrm{d} y^\lambda(\sigma) + ... \right) \delta(\sigma - \sigma^\prime)
\end{align}
with the known tensor hierarchy gauge field strengths \cite{Berman:2020tqn}: 
\begin{align}
F^{(1)}_{\mu \nu} &= 2 \partial_{[\mu_1} A_{\mu_2]} - [ A^{(1)}_{\mu} , A^{(1)}_{\nu} ]_E + \partial A^{(2)}_{\mu \nu}, \label{eq:THFields}\\
F^{(p)}_{\mu_1 ... \mu_{p+1} } &= (p+1) \mathcal{D}_{[\mu_1} A^{(p)}_{\mu_2 ... \mu_{p+1}]} + ... + \partial A^{(p+1)}_{\mu_1 ... \mu_{p+1}} \nonumber
\end{align}
	where $\mathcal{D}_\mu = \partial_\mu - A^{(1)M}_\mu \partial_M$ is associated covariant derivative to $A^{(1)}$ and dots denote some appropriate $\bullet$-products between $A$'s, depending on $p$. Also, as is typical for the relation between current Poisson algebra and antisymmetrised version of the generalised Lie derivative $[ \cdot , \cdot]_E$ some world-volume boundary terms were neglected \cite{Osten:2019ayq,Osten:2021fil}. These field strengths are subject to Bianchi identities of the type
\begin{equation}
D_{[\mu_1} F^{(p)}_{\mu_2 ... \mu_{p+2}]} + ... = \partial F^{(p+1)}_{\mu_1 ... \mu_{p+2}}. \label{eq:THBI}
\end{equation}

\section{End of hierarchy and the compensator field} \label{chap:EndOfHierarchy}
So far, the construction is very general and we barely assumed that there is such a tensor hierarchy structure that is of graded differential Lie algebra type. As 'end of the hierarchy' for the world-volume theories we will understand $\mathcal{R}_{p}$ with $p \geq 9-d$. This is more than is necessary in order to construct the actions of exceptional field theory \cite{deWit:2008gc,Hohm:2013vpa,Hohm:2013uia,Hohm:2014fxa,Berman:2020tqn}, where the construction ends at $\mathcal{R}_{9-d}$. In order to compare to the tensor hierarchy field strengths that are necessary for the construction of the exceptional field theory action, indeed only the field strengths up to $\mathcal{R}_{8-d}$ are relevant.

For the world-volume theory this is not sufficient. For sake of simplicity, let us restrict to the M-theory section. Then, the external space is $($11$-d)$-dimensional. So, in principle one has to consider Poisson brackets \eqref{eq:CurrentTHexternal}
\begin{align}
&{} \quad \{ \pi_\nu (\sigma) , \pi_\mu (\sigma^\prime) \} = - \left( F_{\mu \nu}^{(1) M_1} t^{(1)}_{M_1} + ...  \label{eq:CurrentTHEnd} + F_{\mu \nu \rho_1 .... \rho_{9-d}}^{(10-d) M_{10_d}} t^{(10-d)}_{M_{10-d}} \wedge \mathrm{d} y^{\rho_1} \wedge ... \wedge \mathrm{d} y^{\rho_{9-d}} \right) \delta(\sigma - \sigma^\prime),
\end{align}
so a coupling up to $F^{(10-d)}$. This field strength includes a compensator field $A^{(11-d)}$ from \eqref{eq:THFields}. This indicates that the representations $\mathcal{R}_{10-d}$ and $\mathcal{R}_{11-d}$ are relevant as well. So, in order so that the above construction, \eqref{eq:THmomentum} and \eqref{eq:THFields}, one needs to assume that the representations that were presented for example in table 1 in \cite{West:2018lfn} define a consistent tensor hierarchy, including $\bullet$-products and derivations $\partial$. The cases $d=7,8$ will be discussed explicitly below.

\paragraph{Compensator field.} The details for $\mathcal{R}_{10-d}$ and $\mathcal{R}_{11-d}$ are not relevant in order to compare to expressions of target space exceptional field theory. $\mathcal{R}_{9-d}$ will be slightly modified in comparison to the standard approach, in accordance with \cite{West:2018lfn}. This will introduce the compensator fields at the end of the hierarchy, that are typically associated to redundant degrees of freedom of the dual graviton.

In exceptional field theory, the closure of the gauge algebra relies on a constrained compensator field $\Xi_M \in \mathcal{R}_{8-d} = \bar{\mathcal{R}}_1$, i.e. one with ${\eta_{L_2}}^{M_1 N_1} \Xi_{M_1} \partial_{N_1} = {\eta_{L_2}}^{M_1 N_1} \Xi_{M_1} \Xi_{N_1} = 0$ in addition to the standard section condition \eqref{eq:SectionCondition}. In the world-volume construction, this compensator field will not be a field of $\mathcal{R}_{8-d}$ but of $\mathcal{R}_{9-d}$:
\begin{align}
	\mathcal{R}_{9-d} = \textbf{adj} \quad &\longrightarrow \quad \mathbf{1} \oplus \mathbf{adj}. \label{eq:Representations}
\end{align}
The corresponding tensor hierarchy gauge field $A^{(9-d)}$ will decompose as $(A_{\mathbf{1}}^{(9-d)} , A_{\mathbf{adj}}^{(9-d)})$. The presence of a compensator field at the end of the hierarchy will automatically follow from \eqref{eq:THFields} 
\begin{equation}
\Xi_M = \partial_M A^{(9-d)}_{\mathbf{1}}. \label{eq:NewCompensator}
\end{equation}
This directly implies that, as a gradient, $\Xi_M$ is covariantly constrained. As examples for the above construction, we discuss the cases $d=7,8$ in the following two sections. In comparison to the cases with $d \leq 6$, where the general structure was already discussed in \cite{Osten:2021fil,Osten:2023iwc}. For $d=7,8$, some new aspects appear because the end of hierarchy already appears at $\mathcal{R}_1$, or $\mathcal{R}_2$ respectively.

\section{$E_{7(7)}$ exceptional field theory} \label{chap:E7}
For $E_{7(7)}$ this modification already occurs at $\mathcal{R}_2$. An object in the $\mathcal{R}_2$-representation is then written as $\mathcal{Q}^\mathcal{M} = (\mathcal{Q} , \mathcal{Q}^\alpha ) \in \mathbf{1} \oplus \mathbf{adj}$. The $\eta$-symbol decomposes into the parts
\begin{equation}
	\eta_{\mathcal{L},MN} = (\eta_{\cdot,MN} , \eta_{\alpha,MN} ) = (\Omega_{MN} , (t_{\alpha})_{MN}).
\end{equation}  
Here $(t_\alpha)_{MN}$ are generators of $\mathfrak{e}_{7(7)}$ in the $\mathcal{R}_1$-representation and $\Omega$ is the invariant symplectic form of $E_{7(7)}$. As a consequence, the current algebra \eqref{eq:CurrentAlgebra} now is 
\begin{equation}
	\{z_M (\sigma) , z_N(\sigma^\prime)\} = \left( (t_\alpha)_{MN} \mathcal{Q}^\alpha(\sigma) + \Omega_{MN} \mathcal{Q}(\sigma) \right) \wedge \mathrm{d} \delta(\sigma-\sigma^\prime). \label{eq:E77currentalgebra}
\end{equation}
The $E_{7(7)}$  generalised Lie derivative is $\mathcal{L}_\Lambda \Phi^N = \Lambda^M \partial_M \Phi^N - \Phi^M \partial_M \Lambda^N + {Y^{MN}}_{KL} \partial_M \Lambda^K \Phi^L$ with the $Y$-tensor ${Y^{KL}}_{MN} = - 12 (t^\alpha)^{KL} (t_\alpha)_{MN} + \frac{1}{2} \Omega^{KL} \Omega_{MN}$. The according brane charge conditions take indeed the form of \eqref{eq:BraneChargeConstraint} with the \textit{modified derivation}
\begin{equation}
	(\partial \mathcal{Q})^N = -12 {(t_\alpha})^{MN} \partial_M \mathcal{Q}^\alpha + \frac{1}{2} \Omega^{MN} \partial_M \mathcal{Q} \label{eq:E77ModifiedDerivation}
\end{equation}
Explicitly, the brane charge conditions are
\begin{align}
		\mathcal{Q}^\alpha \wedge \mathrm{d} X^L \partial_L &= - 12 (t^\alpha)^{KL} z_L \partial_K, \qquad \mathcal{Q} \wedge \mathrm{d} X^L \partial_L = \frac{1}{2} \Omega^{KL} z_L \partial_K .  \label{eq:E77ChargeCondition}
\end{align}
Below, it is shown that the Kaluza-Klein monopole is a solution of these conditions with $\mathcal{Q} = 0$ and some non-trivial $\mathcal{Q}^\alpha$.

\paragraph{The external space.} The tensor hierarchy gauge field $A^{(2)} \in \mathcal{R}_2$ decomposes as above \linebreak $A^{(2)}_\mathcal{M} = (A^{(2)}_{\cdot} , A^{(2)}_\alpha)$. In comparison to standard literature on $E_{7(7)}$ \cite{Hohm:2013uia}, the two-form field strength takes the following form in \eqref{eq:THFields}
\begin{align}
	F^{(1) M}_{\mu \nu} &= \partial_\mu A^{(1)M}_{\nu} - \partial_\nu A^{(1)M}_\mu - [A_\mu^{(1)} , A_\nu^{(1)} ]^M_E - 12 (t^\alpha)^{NM} \partial_N A^{(2)}_{\alpha, \mu\nu} + \frac{1}{2} \Omega^{NM} \partial_N A^{(2)}_{\cdot,\mu \nu}.
\end{align}
This and the form in the literature coincide if one identifies the two-form compensator field with the gradient of the $A^{(2)}_\cdot$: $B_{\mu \nu M} = \partial_M A^{(2)}_{\cdot,\mu \nu}$. In order to find a form for the three-form flux $F_{\mu \nu \rho}^{(2) \mathcal{M}} = (H_{\mu \nu \rho}^\alpha , H)$ one would need the derivation $\mathcal{R}_3 \rightarrow \mathcal{R}_2$, where $\mathcal{R}_3 = \mathbf{56} \oplus \mathbf{912}$. With the modified derivation \eqref{eq:E77ModifiedDerivation}, the relevant Bianchi identity between $F^{(1)}$ and the $H$'s takes the canonical form \eqref{eq:THBI}.

\section{$E_{8(8)}$ exceptional field theory} \label{chap:E8}

The relevant representations $\mathcal{R}_1$ and $\mathcal{R}_2$ for $E_{8(8)}$ are both reducible:
\begin{equation}
	\mathcal{R}_1 = \mathbf{1}\oplus \mathbf{adj}, \qquad \mathcal{R}_2 = \mathbf{1} \oplus \mathbf{adj} \oplus \mathbf{3875}.
\end{equation}
Let $K,L,M,...$ denote the adjoint indices. The currents $z_{M_1} \in \mathcal{R}_1$ decompose as $(z,z_M)$, and the $\mathcal{R}_2$-currents as $\mathcal{Q}_{M_2} = ( \mathcal{Q}^{(\mathbf{1})},\mathcal{Q}^{(\mathbf{adj})}_K,\mathcal{Q}_{KL}^{(\mathbf{3875})})$.

 As $E_{8(8)}$-generalisation of \eqref{eq:CurrentAlgebra} the following current algebra is postulated
\begin{align}
	\{ z_M (\sigma) , z_N(\sigma^\prime) \} &= \left( {(\mathbb{P}_{\mathbf{3875}})^{KL}}_{MN} \mathcal{Q}(\sigma)_{KL}^{\mathbf{(3875)}} + {f^K}_{MN} \mathcal{Q}(\sigma)^{\mathbf{(adj)}}_K + \eta_{MN} \mathcal{Q}(\sigma)^{\mathbf{(1)}} \right) \wedge \mathrm{d} \delta (\sigma - \sigma^\prime), \nonumber \\
	\{ z (\sigma) , z_N(\sigma^\prime) \}&= 2 \mathcal{Q}(\sigma)^{(\mathbf{adj})}_N  \wedge \mathrm{d} \delta(\sigma - \sigma^\prime) \label{eq:E88CurrentAlgebra} \\
	\{ z(\sigma) , z(\sigma^\prime) \} &= 0 \nonumber.
\end{align}
The $\eta$-symbol is characterised by the projector $(\mathbb{P}_{\mathbf{3875}})^{KL}_{MN}: \ \mathcal{R}_1 \otimes \mathcal{R}_1 \rightarrow \mathbf{3875}$, the structure constants ${f^K}_{MN}$ of $\mathfrak{e}_{8(8)}$ and the Killing metric $\eta_{MN}$ of $E_{8(8)}$.

A section of these currents in the $\mathcal{R}_1$-representation takes the form
\begin{equation}
	\Phi = (\phi,\psi) = \int \left( \phi^M(\sigma) z_M(\sigma) + \psi(\sigma) z(\sigma) \right) \label{eq:E88Section}
\end{equation}
where also the $\mathbf{1}$ contributes in addition to the $\mathbf{adj}$-representation. It will become clear how this relates to the standard version of the (closing) of the $E_{8(8)}$ generalised Lie derivative. With the brane charge conditions
\begin{align}
	\mathcal{Q}_{MN}^{(\mathbf{3875})} \wedge \mathrm{d}X^K \partial_K &= 14 {(\mathbb{P}_{\mathbf{(3875)}})^{KL}}_{MN} z_L \partial_K \nonumber\\
	\mathcal{Q}^{\mathbf{(adj)}}_M \wedge \mathrm{d} X^K \partial_K &= \frac{1}{2} {f_M}^{KL} z_L \partial_K + z \partial_M \label{eq:E88ChargeCondition} \\
	\mathcal{Q}^{\mathbf{(1)}} \wedge \mathrm{d} X^K \partial_K &= \frac{1}{4} \eta^{KL} z_L \partial_K \nonumber
\end{align}
one obtains for the algebra of the currents \eqref{eq:E88Section}
\begin{align}
	\{\Phi_2, \Phi_1\}^N &= \phi_1^M \partial_M \phi_2^N - \phi_2^M \partial_M \phi_1^N + {Y^{MN}}_{KL} (\partial_M \phi_1^K) \phi_2^L + {f_K}^{MN}( - \partial_M\psi_1 \phi_2^K + \partial_M \psi_2 \phi_1^K ) \nonumber \\
	\{\Phi_2 , \Phi_1 \}^\cdot &= \phi_1^M \partial_M \psi_2 - \phi_2^M \partial_M \psi_1 + (\partial_M \phi_1^K) \phi_2^L {f^M}_{KL} + ( - \partial_M \psi_1 \phi_2^M + \partial_M \psi_2 \phi_1^M) \label{eq:E88Closure}
\end{align}
with the $E_{8(8)}$ $Y$-tensor ${Y^{MN}}_{KL} = 14 {(\mathbb{P}_{\mathbf{3875}})^{KL}}_{MN} + \frac{1}{2} {f_P}^{MN} {f^P}_{KL} + \frac{1}{4} \eta^{MN} \eta_{KL}$. The first line of \eqref{eq:E88Closure} relates easily to the more standard commutator of two generalised Lie derivatives \linebreak $[\mathcal{L}_{(\phi_1^M,\Sigma_{1M})} ,\mathcal{L}_{(\phi_2^M,\Sigma_{2M})} ] $ in \cite{Hohm:2014fxa} with the identification
\begin{equation}
\Sigma_M = \partial_M \psi.
\end{equation}
So, again as discussed in section \ref{chap:EndOfHierarchy} the compensator field is introduced here as the gradient of the singlet part of \eqref{eq:E88Section}.

\paragraph{External space and tensor hierarchy gauge field.} The complete external momentum is given by
\begin{align}
\pi_\mu = p_\mu - A_\mu^M z_M - B_\mu z - C^{KL}_{\mu \nu} \mathcal{Q}^{\mathbf{(3875)}}_{KL} \wedge \mathrm{d} y^\nu - C^K_{\mu \nu} \mathcal{Q}_K^{\mathbf{(adj)}} \wedge \mathrm{d} y^\nu - C_{\mu \nu} \mathcal{Q}^{\mathbf{(1)}} \wedge \mathrm{d} y^\nu \label{eq:E88ExtMomentum}
\end{align}
up to the $\mathcal{R}_{11-d} = \mathcal{R}_3$-fields that are irrelevant in order to compare to target space exceptional field theory. Again, the dynamics in a complete supergravity background, including the external directions are characterised by the Hamiltonian \eqref{eq:Hamiltonian}, the current algebra \eqref{eq:E88CurrentAlgebra} and
\begin{equation}
\{\pi_\nu(\sigma) , \pi_\mu(\sigma^\prime)\} = - \left(F_{\mu \nu}^M z_M(\sigma) + G_{\mu \nu} z(\sigma) + H_{\mu\nu\rho}^{M_2} \mathcal{Q}_{M_2}(\sigma) \right) \delta(\sigma - \sigma^\prime)
\end{equation}
with the $E_{d(d)}$-covariant field strengths
\begin{align}
	F_{\mu \nu}^N &= 2 \partial_{[\mu} A_{\nu]}^N +  {Y^{MN}}_{KL} A^K_{[\mu} \partial_M A^L_{\nu]} +  14 {(\mathbb{P}_{\mathbf{3875}})^{MN}}_{KL} \partial_M C^{KL}_{\mu \nu} + {f_L}^{MN} \left( A_{[\mu}^L \partial_M B_{\nu]} + \frac{1}{2} \partial_M C_{\mu\nu}^L \right) + \frac{1}{4} \eta^{MN} \partial_M C_{\mu \nu} \nonumber \\
	G_{\mu \nu} &= 2 D_{[\mu} B_{\nu]} + {A_{[\mu}^K} \partial_M {A_{\nu]}^L} {f^M}_{KL} \label{eq:E88Fieldstrength}
\end{align}
which can be computed from \eqref{eq:E88ExtMomentum}

In order to obtain the expression for $F_{\mu\nu}^M$ from the standard exceptional field theory literature, the following field redefinition has to be made: ${C^\prime_{\mu \nu N}}^K = \frac{1}{2} A^K_{[\mu} \partial_N B_{\nu]} + \frac{1}{4} \partial_N C_{\mu\nu}^K$. Again, ${C^\prime_{\mu \nu N}}^K$ is covariantly constrained.

As in the second line in \eqref{eq:E88Closure}, the $G_{\mu\nu} \in \mathbf{1}$ field strength does not seem to be related in an obvious way to the quantities in standard $E_{8(8)}$ exceptional field theory \cite{Hohm:2014fxa}, where one would expect a quantity like $G^\prime_M = \partial_M G$. Let us note that in addition to certain trivial gauge parameters that appear in $E_{8(8)}$ exceptional field theory, there is the freedom to add world-volume boundary terms for the world-volume theories.

\section{Example: the KK-monopole world-volume} \label{chap:KKM}
As a novel example for the canonical exceptional structure, the Kaluza-Klein monopole (KKM) world-volume theory is presented. As will be shown, this requires $E_{d(d)}$ generalised geometry with $d \geq 7$. This is due to the fact that despite it being only a 6-brane it requires an additional isometry direction perpendicular to the world-volume. 

\paragraph{Hamiltonian dynamics from the gauged $\sigma$-model.} Following \cite{Bergshoeff:1997gy} the world-volume dynamics in 11-dimensional supergravity can be described as a gauged $\sigma$-model. The U$(1)$-isometry, corresponding to the special isometry direction of the supergravity solution, is gauged. With conventions, in which $\underline{\alpha} , \underline{\beta},... = 0,...,6$ and $\alpha,\beta,...=1,...6$, the Polyakov-type action of the KKM coupling to a background metric $g_{mn}$ is:
\begin{equation}
	S = - \frac{T_{KK}}{2} \int \mathrm{d}^7 \sigma \ \sqrt{-h} \left( k^{4/7} h^{\underline{\alpha} \underline{\beta}} D_{\underline{\alpha}} x^m D_{\underline{\beta}} x^n g_{mn} - 5 \right) \nonumber
\end{equation}
$k^2 = - k^m k^n g_{mn}$ is the norm of a Killing vector $k^m (x)$. This Killing vector is associated to the special isometry direction of the KKM-supergravity solution. $D_\alpha x^m = \partial_\alpha x^m + C_\alpha k^m$ is a covariant derivative, with auxiliary world-volume 1-forms $C_\alpha$. The equations of motion of $C_\alpha$ are constraints that ensure that the world-volume is orthogonal to the Killing vector
\begin{equation}
\Phi^{\underline{\alpha}} = - T_{KK} \sqrt{-h} h^{\underline{\alpha} \underline{\beta}}  k^{4/7} D_{\underline{\beta}} x^m k^n g_{mn} \approx 0. \label{eq:KillingConstraints} 
\end{equation}
In particular $\Phi^0 \sim p_m k^m$. The involution of these constraints is non-trivial -- in particular \linebreak $\{\Phi^0(\sigma) , \Phi^\alpha (\sigma^\prime) \}$ -- but can be understood as the consequence of the natural condition on the world-volume\footnote{In target coordinates associated to the isometry, this means that the coordinate field does not appear in the action, as observed.} 
\begin{equation}
k^m \partial_m f(x (\sigma)) = 0. \label{eq:KillingWorldVolume}
\end{equation}
The equations of motion for the world-volume metric $h_{\underline{a} \underline{\beta}}$ are
\begin{equation}
h_{\underline{\alpha} \underline{\beta}} = k^{4/7} D_{\underline{\alpha}} x^m D_{\underline{\beta}} x^n g_{mn}.
\end{equation}
With that the Hamiltonian and the spatial diffeomorphism constraints can be computed as:
\begin{align}
	H &\approx \frac{1}{2} \left( g^{mn} p_m p_n + k^4 D_{\alpha_1} x^{m_1} ... D_{\alpha_6} x^{m_6} D_{\beta_1} x^{n_1} ... D_{\beta_6} x^{n_6} \epsilon^{\alpha_1 ... \alpha_6} \epsilon^{\beta_1 ... \beta_6} g_{m_1 m_1} ... g_{m_6 n_6} \right), \label{eq:KKMHamiltonian} \\
	P_\alpha &\approx p_\mu \partial_\alpha x^\mu \approx 0  \nonumber
\end{align}
after taking $T_{KK} = 1$, choosing the gauge $h^{00} = h^{-1/2} k^{-4/7}$ and $h^{0\alpha} = 0$, and by virtue of the constraints \eqref{eq:KillingConstraints}.

\paragraph{Realisation of the Hamiltonian via $E_{d(d)}$-currents.} Let us postulate the following currents
\begin{align}
z_M &= ( z_m , z^{m_1 m_2} , z^{m_1 ... m_5} , z^{m_1 ... m_7,m } ) \label{eq:KKMCurrentSimple}= (p_m , 0 , 0 , k^m k^{[m_1} \mathrm{d} x^{m_2} \wedge ... \wedge \mathrm{d} x^{m_7]}).
\end{align}
As for the supergravity solution, the existence of the Killing vector $k$ is crucial. For the current algebra approach, one needs the Killing vector $k^\mu$ for the possibility to have a spatial $6$-form that fits into the mixed symmetry tensor $z^{m,m_1...m_7}$. The Hamiltonian \eqref{eq:KKMHamiltonian} can indeed be reproduced by a Hamiltonian \linebreak $H = \frac{1}{2} \int \mathcal{H}^{MN} z_M z_N$ by assuming that the relevant components of the generalised metric are
\begin{equation}
\mathcal{H}^{mn} = g^{mn}, \ \mathcal{H}_{m,m_1 ... m_7; n,n_1...n_7} = g_{mn} g_{m_1 [\underline{n_1}} ... g_{m_7 \underline{m_7}]} . \label{eq:KKMGravGenMetric}
\end{equation}
For this it is crucial to employ the constraints \eqref{eq:KillingConstraints} and notice that $z^{m,m_1 ... m_7}$ can also be expressed as $k^m k^{[m_1} D x^{m_2} \wedge ... \wedge D x^{m_7]}$. Like this the dynamics can be easily generalised to arbitrary backgrounds by taking other generalised metrics apart from the geometric pure metric one \eqref{eq:KKMGravGenMetric}. The spatial diffeomorphism constraints from \eqref{eq:KKMHamiltonian} imply also the conjectured canonical $E_{d(d)}$-invariant form \eqref{eq:Hamiltonian}.

\paragraph{The $E_{d(d)}$-covariant current algebra.}
Let us choose target space coordinates $x^m = (x^{\bar{m}},w)$ that are adapted to the isometry where $w$ is the coordinate associated to the Killing vector $k$, i.e. $k = \partial_w$.\footnote{When staying in general coordinates one will obtain a twist, similar to those obtained in a different frame considered in \cite{Osten:2021fil}. The result would still exhibit $E_{d(d)}$-covariance but not in the canonical form \eqref{eq:CurrentAlgebra}.}. The constraints \eqref{eq:KillingConstraints} and \eqref{eq:KillingWorldVolume} then imply that $\partial_w f = 0$ for any physical function $f$ on the KKM world-volume and $p_w = 0$. In these coordinates, the only non-vanishing components of $z_M$ \eqref{eq:KKMCurrentSimple} are 
\begin{equation}
	z_{\bar{m}} = p_{\bar{m}}, \qquad z^{w,\bar{m}_1 ... \bar{m}_6 w} = \mathrm{d}x^{\bar{m}_1} \wedge ... \wedge \mathrm{d} x^{\bar{m}_6} \label{eq:KKMcurrentSimpleCoord}
\end{equation}
with $\{ z_M(\sigma) , z_N(\sigma^\prime) \} = \eta_{\mathcal{P} MN} \mathcal{Q}^{\mathcal{P}} \wedge \mathrm{d} \delta(\sigma-\sigma^\prime)$  for $\mathcal{Q}^{w,\bar{m}_1....\bar{m}_5 w} = \mathrm{d}x^{\bar{m}_1} \wedge ... \wedge \mathrm{d} x^{\bar{m}_5}$. 

Let us now specify to the context of $E_{7(7)}$, where the $\eta$-symbols are introduced in section \ref{chap:E7} and their explicit GL$(d)$-decomposition can be found for example in \cite{Sakatani:2017xcn}. For \eqref{eq:KKMcurrentSimpleCoord} together with $\partial_w f = 0$, implies that $\Omega^{KL} z_K \partial_L f = 0$. Hence, the second part of the $E_{7(7)}$ $Y$-tensor ${Y^{KL}}_{MN} =  \eta^{\mathcal{P}KL} \eta_{\mathcal{P}MN} + \frac{1}{2} \Omega^{KL} \Omega_{MN}$ does not contribute and the generalised Lie derivative \eqref{eq:GeneralisedLieViaCurrent} is reproduced on the KKM phase space.

%
%
%
%
%
%
%
%

\section{Outlook}

This article explored a unifying Hamiltonian formulation of brane dynamics in string and M-theory. This formulation is a generalisation of the so-called $\mathcal{E}$-model for two-dimensional non-linear $\sigma$-models. These two-dimensional models are particularly rich of additional structures. Recently, such Hamiltonian formulation has been applied to the bosonic part of the heterotic string \cite{Hatsuda:2022zpi} in order to explore new integrable models \cite{Osten:2023cza}. Similar successes might be feasible for the exceptional string dynamics \cite{Arvanitakis:2018hfn}, i.e. the $(p,q)$-string.  

A second point, where this approach is expected to give new results, lies in the dynamics and nature of exotic branes \cite{deBoer:2010ud,deBoer:2012ma,Bakhmatov:2017les,Berman:2018okd}. In order to obtain insights into the exotic brane landscape with the present approach the following issues still need to be clarified:
\begin{itemize}
\item For branes of arbitrary spatial extension and amount of 'special' or 'smeared' external dimensions, extensions of the setup of this paper to $E_{d(d)}$ with $d \leq 11$ are needed. Actions for exotic five-branes, to which one could compare the present approach, were given in \cite{Kimura:2014upa,Kimura:2016anf}.

For the KK monopole these smeared directions corresponded to a gauge symmetry of the world-volume model. It is expected that this would be true generally. Hence, also an extension of exceptional generalised geometry that accounts for additional gauge symmetries analogous to \cite{Polacek:2013nla} is necessary. There has been a recent approach to this task in \cite{Sakatani:2022auu}. This is based on the construction of actions \cite{Sakatani:2016sko,Sakatani:2017vbd,Sakatani:2020umt}. Its relationship to the approach based on current algebra in this letter should be clarified.

\item In order to solve the brane charge constraints \eqref{eq:BraneChargeConstraint}, the algebraic structure of the tensor hierarchies up to $\mathcal{R}_{11-d}$ should be known. Explicit expressions for the $\eta$- and $D$-symbols \eqref{eq:TensorHierarchy} in both the M-theory and type IIb decomposition would be particularly necessary.

\item The tensor hierarchy current algebra and in particular even the $z$-$z$ current algebra for the $E_{7(7)}$- and $E_{8(8)}$-theories are in general not skewsymmetric. 

In fact, for all the brane models encountered so far, the symmetric part drops out. It is possible that this pattern continues for the exotic branes. But it might also be possible that these 'wrong statistic' Poisson brackets indicate some ghost fields associated to some additional gauge symmetry of the exotic brane world volume theories.
\end{itemize}
Also, the relation to other approaches to unified formulations of world-volume theories in string and M-theory -- in particular \cite{West:2018lfn} and recently \cite{Hatsuda:2023dwx} -- should be made. In former, a linearised version of brane dynamics was derived in the setting of $E_{11}$ exceptional field theory. A connection to the approach in this article seems evident and should be explored.

\section*{Acknowledgements}
The author thanks Alex Swash for comments on the draft and Chris Blair for discussions about related topics.

This research is part of the project No. 2022/45/P/ST2/03995 co-funded by the National Science Centre and the European Union’s Horizon 2020 research and innovation programme under the Marie Sk\l odowska-Curie grant agreement no. 945339.

\vspace*{10pt}
\includegraphics[width = 0.09 \textwidth]{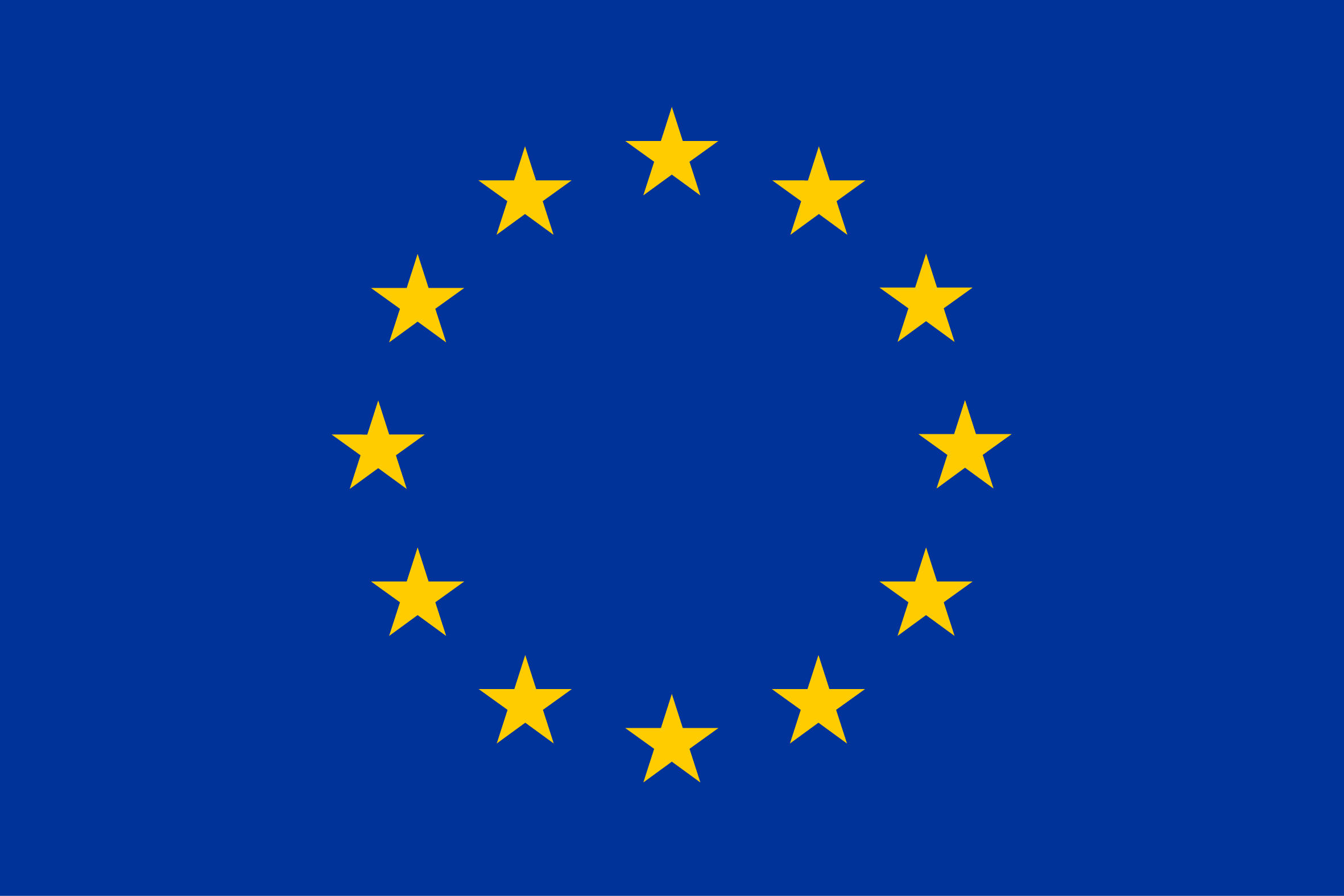} $\quad$
\includegraphics[width = 0.7 \textwidth]{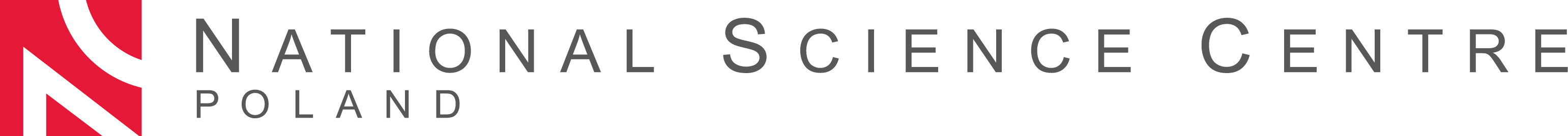}

\bibliographystyle{jhep}
\bibliography{References}
\end{document}